\documentclass[conference]{IEEEtran}
\IEEEoverridecommandlockouts
\usepackage{cite}
\usepackage{amsmath,amssymb,amsfonts}
\usepackage{algorithmic}
\usepackage{graphicx}
\usepackage{textcomp}
\usepackage{xcolor}
\usepackage{soul}
\setstcolor{red}
\usepackage{titlesec}
\usepackage{subcaption}

\def\BibTeX{{\rm B\kern-.05em{\sc i\kern-.025em b}\kern-.08em
    T\kern-.1667em\lower.7ex\hbox{E}\kern-.125emX}}
\begin{document}

\title{Distributed MIMO Precoding with Routing Constraints in Segmented Fronthaul \\

}

\author{\IEEEauthorblockN{1\textsuperscript{st} Jale Sadreddini}
\IEEEauthorblockA{\textit{Ericsson Canada} \\
Ottawa, Canada \\
jale.sadreddini@ericsson.com}
\and
\IEEEauthorblockN{2\textsuperscript{nd} Omer Haliloglu}
\IEEEauthorblockA{\textit{Ericsson Research} \\
Istanbul, Turkiye\\
omer.haliloglu@ericsson.com}
\and
\IEEEauthorblockN{3\textsuperscript{rd} Andres Reial}
\IEEEauthorblockA{\textit{Ericsson Research} \\
Lund, Sweden \\
andres.reial@ericsson.com}
}
\maketitle

\begin{abstract}
Distributed Multiple-Input and Multiple-Output (D-MIMO) is envisioned to play a significant role in future wireless communication systems as an effective means to improve coverage and capacity. In this paper, we have studied the impact of a practical two-level data routing scheme on radio performance in a downlink D-MIMO scenario with segmented fronthaul. At the first level, a Distributed Unit (DU) is connected to the Aggregating Radio Units (ARUs) that behave as cluster heads for the selected serving RU groups. At the second level, the selected ARUs connect with the additional serving RUs. At each route discovery level, RUs and/or ARUs share information with each other. The aim of the proposed framework is to efficiently select serving RUs and ARUs so that the practical data routing impact for each User Equipment (UE) connection is minimal. The resulting post-routing Signal-to-Interference plus Noise Ratio (SINR) among all UEs is analyzed after the routing constraints have been applied. The results show that limited fronthaul segment capacity causes connection failures with the serving RUs of individual UEs, especially when long routing path lengths are required. Depending on whether the failures occur at the first or the second routing level, a UE may be dropped or its SINR may be reduced. To minimize the DU-ARU connection failures, the segment capacity of the segments closest to the DU is set as double as the remaining segments. When the number of active co-scheduled UEs is kept low enough, practical segment capacities suffice to achieve a zero UE dropping rate. Besides, the proper choice of maximum path length setting should take into account segment capacity and its utilization due to the relation between the two.

\end{abstract}


\section{Introduction}
Distributed MIMO (D-MIMO), also known as "cell-free massive MIMO" is a key technology candidate for the 6G. The basic idea is to distribute the antennas geographically but make them operate together \cite{R1}. 
Antenna distribution provides macro diversity which significantly reduces signal blocking that is important for high-frequency bands and provides a more uniform User Equipment (UE) performance by limiting worst-case path loss. Joint transmission schemes provide high performance with high reliability which is essential for critical communications, e.g., industrial scenarios. To make deployment of a large number of Radio Units (RUs) simple and cost-efficient, various solutions have been proposed, such as Radio Stripes \cite{R2} and Radio Weaves \cite{R3}, which have a common feature that they use a shared Fronthaul (FH) together with a high degree of integration and miniaturization. The FH structure used by e.g., Radio Stripes and Radio Weaves may be referred to as segmented FH such that each RU is connected to one or more neighboring RUs via interfacing segments that can be used for transferring power, Downlink (DL) data packets and precoding weights, Uplink (UL) combining weights and symbol estimates, etc. An important property of such FH structure is that a given RU is generally not directly connected to the Distributed Unit (DU) of D-MIMO but signals to/from it need to pass multiple segments to reach their destinations. To ensure reliable and efficient communication between multiple UEs and RUs, careful routing solutions are required to efficiently utilize the data transfer capabilities of the segments. 

In a UE-centric approach in D-MIMO, RU grouping for each UE is performed to define the serving RU subset that jointly transmits data to the UE. The grouping may be based on radio criteria, e.g., channel qualities, the maximum number of UEs that an RU can simultaneously serve, etc. Per-group precoding/combining can be done, either in the DU or in an Aggregating Radio Unit (ARU), which is the responsible RU in the group that can perform computations on behalf of the serving RUs and convey group information to the DU. A routing solution is then determined to exchange data between the DU and RUs. If a routing path for a certain RU cannot be established due to FH segment capacity limitations, the resulting joint transmission may be highly sub-optimal since a missing transmission component will distort the spatial energy focusing, interference null steering, or resulting UL combining, and may lose diversity gain. To improve robustness, joint grouping and routing approaches may be envisioned that explicitly test different grouping options and check their resulting routing solutions to determine the best grouping options. However, such solutions would be feasible only for a very low number of UEs and RUs; their complexity for scenarios of practical interest is prohibitive. 

\subsection{Related Works}
D-MIMO with segmented FH was proposed as a scalable approach, using a compute-forward architecture where FH links cascade interconnecting a DU with multiple RUs \cite{R2}. The majority of works on D-MIMO have considered coherent DL data transmission in scenarios with unlimited FH, such as in \cite{R8,R9,R10,R11}. The works in \cite{R5,R6,R7} have investigated the impact of the limited capacity of FH networks on the UL performance of D-MIMO systems. In \cite{R12}, the authors proposed a transmission probability-based RU clustering scheme in user-centric D-MIMO to reduce the FH capacity requirement. In \cite{R13}, authors compare the performance of distributed and centralized precoding with limited FH capacity in DL D-MIMO setup. 
Authors in \cite{R15} consider a point-to-multipoint FH topology where an RU subset shares a serial FH link with limited capacity and develops iterative power control and RU scheduling algorithm. \cite{R16} analyzes the impact of individual and cumulative failures on RUs and FH segments in D-MIMO with segmented FH using Markov chains. 


\subsection{Perspective of this work}
Although various FH topology options have been investigated, to the best of our knowledge, there are currently no examples of implementing a routing algorithm in any D-MIMO topology to analyze the performance of DL D-MIMO with routing constraints in segmented FH. In this work, we pursue a low-complexity approach to jointly consider RU grouping and routing algorithms, followed by a precoding method and eventually analyze the performance of DL D-MIMO in the segmented FH with routing constraints, e.g., FH segment capacity and path length. The overall aim is to keep the  practical routing impact on radio performance low. A two-level routing-based DL transmission is implemented where, in L1, the DU is connected to the ARU. In L2, the selected ARUs will be connected to the remaining serving RUs in the group. A route discovery procedure based on the strategy of the previously proposed Hybrid Multi-Path Routing algorithm (HyMPRo) defined in \cite{R18,R19}, has been applied at each routing level to find route alternatives and determine the best ones based on the segment utilization and path length. Although we consider a mesh network with grid topology, the proposed two-level routing principles can also be applied to any kind of topology

The rest of the paper is organized as follows. In Section II, we describe the D-MIMO system model and segmented FH. In section III, we provide numerical results and discussions. Finally, Section IV concludes the paper.

\section{System Model}
\begin{figure}[tb!]
\centerline{\includegraphics[width=0.45\textwidth]{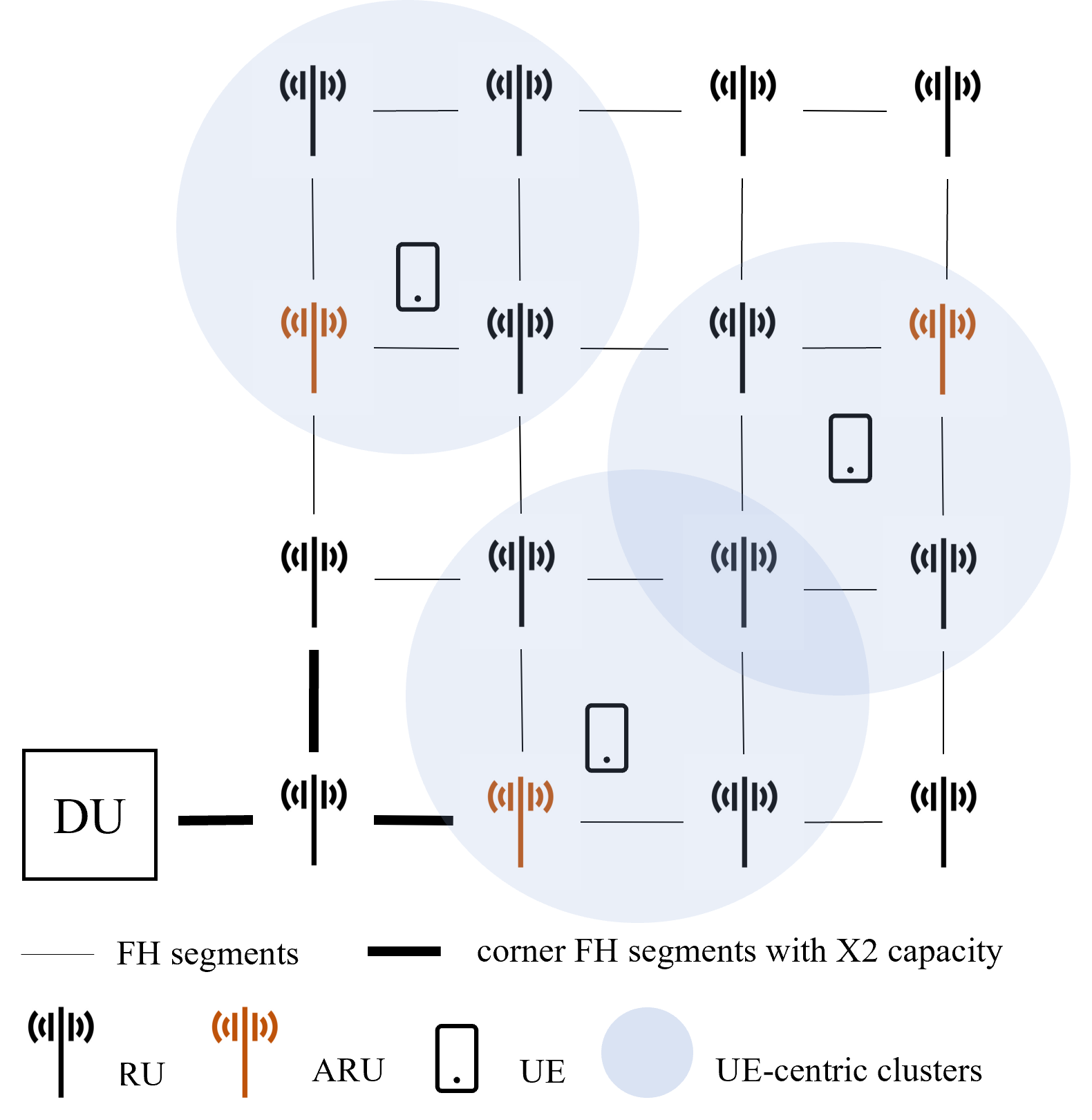}}
\caption{Example D-MIMO network with segmented FH.}
\label{fig:system_model}
\end{figure}

The considered network consists of a DU, \emph N geographically distributed RUs, and \emph K UEs, shown in Fig. \ref{fig:system_model}, where there are \emph S segments that are defined as FH links (e.g., fiber-optic cables) between two RUs. Each RU is connected to its neighbors, and generally not directly to the DU, via segments. The \emph k-th UE will connect to $M_k$ RUs where $M_k$ is the number of RUs in the serving subset for the \emph k-th UE which can be determined by the DU. It is assumed that RUs are equipped with \emph T antennas and UEs have a single antenna. (It is straightforward to generalize the mathematical framework suitably for multi-antenna UEs.) The information-bearing symbol intended for the \emph k-th UE is denoted by $q_k$, where E$\{q_k q_j^* \}= \delta _{jk}$, as $\delta _{jk}=1$ for $j=k$, $\delta _{jk}=0$ otherwise. $h_{kn} \in C^{1 \times T}$ represent the complex-valued channel coefficients between the \emph n-th RU and \emph k-th UE. Each RU has a maximum transmission power limit represented by \emph P, and zero-mean circularly symmetric additive white Gaussian noise with variance $ \sigma_{\eta}^{2}$ is denoted by $\eta_k$. Then the received signal for the \emph k-th UE can be expressed as
\begin{equation}%
	\begin{array}
		[l]{ll}%
		\mathbf{y}_k(\boldsymbol{\omega})=
		\sum_{n=1}^{M_k}\sqrt{P_n}\mathbf{h}_{kn}\boldsymbol{\omega}_{kn}\mathbf{q}_k+ \\
		\sum_{j=1,j\neq k}^{K} \sum_{n=1}^{M_j}\sqrt{P_n}\mathbf{h}_{jn}\boldsymbol{\omega}_{jn}\mathbf{q}_j+\eta_k, \forall k,
	\end{array}
	\label{eqn:receivedsignal}%
\end{equation}
where the first term is the intended signal received from the $M_k$ RUs, and the following ones represent the interference due to multi-user transmission and noise term, respectively. In case of a perfect FH network, SINR ($\gamma$) for the \emph k-th UE is given as follows:
\begin{equation}
	\gamma_k(\boldsymbol{\omega})=\frac{\lvert \sum_{n=1}^{M_k}\sqrt{P_n}\mathbf{h}_{kn}\boldsymbol{\omega}_{kn} \rvert^2}
	{\sum_{j=1,j\neq k}^{K} \lvert \sum_{n=1}^{M_j}\sqrt{P_n}\mathbf{h}_{jn}\boldsymbol{\omega}_{jn} \rvert^2 + \sigma_{\eta}^2}, \forall k.
	\label{eqn:SINR}%
\end{equation}

\subsection{Distributed MIMO over segmented fronthaul}
In the given segmented FH structure, each RU is connected to one or more neighboring RUs via interfacing segments, where the communication between an RU and the DU can be realized over multiple segments. Fig. \ref{fig:system_model} depicts a system configuration where each UE is assigned  multiple serving RUs, e.g., $M=5$, where one of them is chosen as an ARU to receive the UE data packets from the DU and distribute them to the remaining serving RUs. It should be noted that since all data will be passed from DU to ARUs, the closest segments to the DU will be occupied first. Thus, to avoid L1 failures due to a few critical segments, the capacity of the corner segments is doubled. 

\subsection{Subset Selection and Routing Framework}
\begin{figure}[t!]
\centerline{\includegraphics[width=0.5\textwidth]{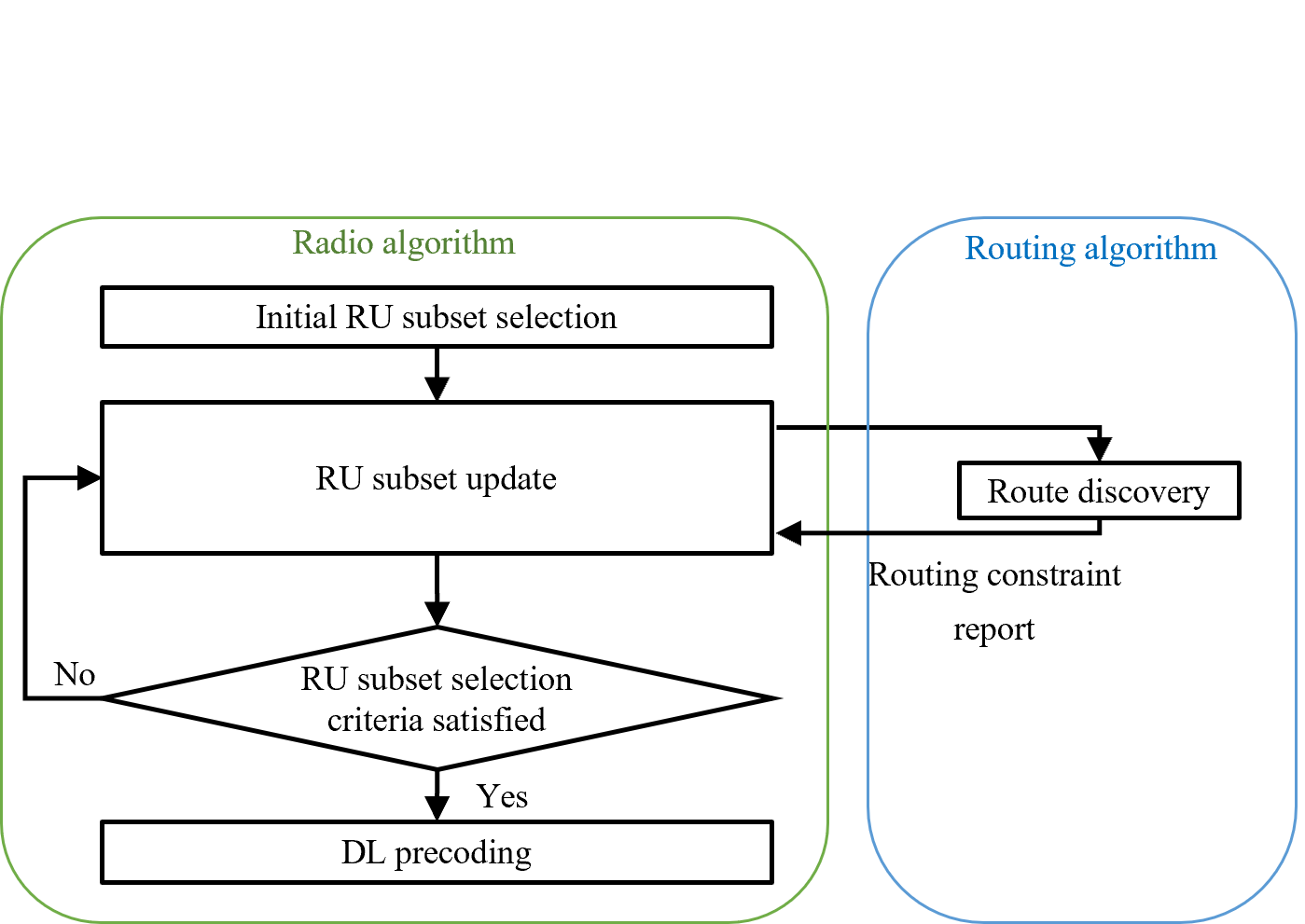}}
\caption{Structure of the proposed model.}
\label{fig:proposed_model}
\end{figure}

Our processing sequence can be captured in four steps: 1) initial per-UE RU subsets selection, 2) routing 3) RU subsets update, and 4) precoding. The radio algorithm, which includes serving RU subset selection and precoding, provides an initial list of RUs to the routing algorithm. The routing algorithm determines packet routing paths from the source node, i.e., DU, to the destination nodes, i.e., serving RU(s) via Route Request (RREQ) packet \cite{R18}, where each RU reports back the associated success and/or failure, indicating whether they could serve the UEs. Then, the radio algorithm uses the routing report to modify the UE-centric RU subsets and computes the precoding weights for the updated RU subsets. It should be noted that the proposed steps can be run in several iterations until meeting the RU subset selection criteria, e.g., minimum SINR and a minimum number of serving RUs per UE, etc. The structure of the proposed sequence is depicted in Fig. \ref{fig:proposed_model}.

\subsubsection{RU subset selection}
UE-centric RU selection has been used where the subset of RUs serving each UE is selected to maximize link quality, e.g. to provide a sufficiently high SINR. Such subsets may not be disjoint but may partially overlap with neighboring subsets serving other UEs. The RUs have been grouped in a way that their large-scale fading coefficients contribute at least $\alpha\%$, e.g., $95\%$, towards the UE such that $\Sigma_{m=1}^{M_k}\{{h_{kn}^{'}⁄\Sigma_{m^{'}}^{N}{h_{k m^{'}}^{'} }}\} \geq \alpha\%$, where $\{{h_{k1^{'}},h_{k2^{'}},…,h_{kN^{'}} }\}$ is the set of large scale fading coefficients, i.e., $h_{kn}^{'}=  \mathbb{E} \{ {|h_{kn}|^{2} } \}$ \cite{R20}.
For the sake of simplicity, the initial subsets selected to serve the $k$-th UE, consists of $M_k$ RUs which are the RUs having the best channel gains with the $k$-th UE, and this subset is denoted by $RU^k$.
RU-UE association matrix in general is denoted by $\mathbf{A}$, having binary-valued elements, i.e., $a_{nk}=\{0,1\}$ meaning that $a_{nk}=1$ if $n$-th RU serves to $k$-th UE, zero otherwise.

\subsubsection{Routing}
A two-level routing procedure is proposed, in which L1 generates the route from DU to ARU$_k$, and L2 discovers route(s) from the ARU$_k$, to the other serving RUs of $k$-th UE. We use the HyMPRo routing algorithm \cite{R18}. Before starting route discovery, routing constraints, e.g., FH segment capacity and maximum path length, are determined for the multi-path route discovery utility function which is found as follows:

\textbf {FH segment capacity}: The occupancy rate of the FH segment capacity can be defined as
 \begin{equation} \label{Eq3bb}
    \beta\left(S_z\right)=UsedCap(S_z)/TotalCap(S_z),
\end{equation}
where $UsedCap(S_z)$ is the current count of packets in the segment and $TotalCap(S_z)$ is the maximum number of packets that can be conveyed in the \emph{z}-th segment where $z=1..Z$, $Z$ is the total number of segments in the network topology. The utilizable rate of the total FH segment capacity for each route, $\rho_{Ut}$, is evaluated at the DU as
 \begin{equation} \label{Eq4aa}
  \rho_{Ut} (R)=\Pi_{g=1}^{G}\left(1- \beta(S_z )\right),z\in Z,   
\end{equation}
where $\rho_{U_t} (R)$ can be considered as a congestion awareness metric where $0 \leq \beta(S_z ) < 1$. \emph{G} is the total number of the involved segments at each route \emph{R} during the route discovery process. 
The rate of utilizable segment capacity of a potential route becomes critical if the data traffic increases. Thus, it is more rational to forward the data in a newly generated route to circumvent the highly utilized segments. Otherwise, data packages of the next route run the risk of being dropped in a region of high traffic load.

\textbf {Path Length}: Apart from the segment capacity, the number of segments that can be involved in a route plays a major role in successful connection vs. dropping. The path length, \textit{L(R)}, is the number of segments in route R, where the maximum path length sets a limit on the number of segments.  If $\beta(S_z )\cong 1$, the segment is almost congested. That is why a higher maximum path length value relaxes the problem of finding alternative transmission routes. Although it increases computational complexity and may increase communication latency, the UE can still be served. Considering both constraints, the utility function can be calculated as follows:
 \begin{equation} \label{Eq5}
   f_ {\left(R\right)}=\rho_{Ut} (R)/L(R).
\end{equation}

After all instances of the sent RREQ, with different piggybacked information depending on the followed route, arrive at the destination, the utility function values are evaluated for each candidate route, $R_1, ..., R_{\Psi}$, where $\Psi$ is the total number of candidate routes, $0 \leq \Psi \leq L$. Among the candidate routes, the DU chooses the best route, \textit{BR}, given as follows:
\begin{equation} \label{Eq6}
    BR=\begin{cases} 
        \text{argmax}\{ f(R_{nr})\}, & 1 \leq nr \leq \Psi , \text{ if } \Psi \geq 1, \\
        0, & \text{otherwise}.
    \end{cases} 
\end{equation}

\subsubsection{RU subset Updates}
RU subsets will be modified in the radio algorithm based on the routing outcome information subject to practical routing limitations in the FH network. In our solution, only successfully connected initial serving RUs are kept as part of the modified serving RU set for a given UE.
Any RUs in the initial subset of RUs, serving a UE, that could not be connected after the routing step, will be excluded from the serving set. 
Thus, $BR_{n,k}=0$, means that the route discovery algorithm, considering the utilizable segment capacity of each segment connected to the RUs in $RU^k$ and path length criteria, can not find any suitable route from ARU$_k$ to RU$_{n} \in RU^k$. Therefore, RU subsets for all UEs need to be updated as follows:
\begin{equation} \label{8}
    \mathrm{A}_{n,k}=\begin{cases} 
        0, & \text{if } BR_{n,k}=0, \\
        1, & \text{otherwise}.
    \end{cases}
\end{equation}

If any of the initially selected \emph{K} ARUs serving a UE could not connect, the UE will not get service and consequently will be dropped, hence $k$-th column of $\mathbf{A}$ will be updated as zero.

\subsubsection{Precoding}
Proper precoding in interference-limited scenarios ensures constructive addition of the intended signals at the location of corresponding UEs and preferably also inter-user interference suppression. Precoding can be carried out in the central entity, i.e., DU using the information gathered from all RUs, semi-distributed in the ARUs, or locally at each RU. 
In our setup, Centralized Zero-Forcing (CZF) precoding is carried out in the DU. The local data at the DU needs to be conveyed to the first ARU, then RUs, using the routing information obtained in the routing discovery step. 


The precoding matrix, which is denoted by $\mathbf{W} \in C^{NT \times K}$ can be found via pseudo inverse of the effective channel matrix ($\mathbf{H}_{eff}$), such that $\mathbf{W}=\mathbf{H}_{eff}^{\dagger}$ where $\mathbf{H}_{eff}=\mathbf{H} \cdot \mathbf{A}$, where $\cdot$ is the dot product operation, $\mathbf{H} \in C^{NT \times K}$ is the channel gain matrix and given by $\mathbf{H}=[\mathbf{H}_{1}^{H} \: \mathbf{H}_{2}^{H} … \mathbf{H}_{K}^{H}]^{H}$, in which $(.)^{H}$ is the conjugate transpose operation and $\mathbf{H}=[\mathbf{h}_{k1} \: \mathbf{h}_{k2} … \mathbf{h}_{kN}]$ is the channel gain matrix for transmission to the \emph{k}-th UE. When the precoding coefficients are determined, the routing algorithm for packet forwarding is carried out in the network, e.g., between the DU and/or individual ARUs that execute the routing based on the updated RU subsets.

\section{	Performance Evaluation}
In this section, we discuss example cases in which the number of UEs and FH segment capacity strongly affect the successful connection ratio and the resulting radio performance. Only one iteration has been run between radio and routing algorithms to limit complexity, additional iterations may further improve the performance.
\subsection{	Simulation parameters and assumptions}
The performance of the large-scale D-MIMO system depicted in Fig.  \ref{fig:system_model} is analyzed in a DL indoor factory scenario over an area of 100m × 100m with randomly distributed 1000 blockers. The performance has been evaluated for $N=16$ regularly deployed RUs, and $K=\{8, 15\}$ uniformly distributed UEs. It is assumed that UEs and RUs are stationary, and initial serving subsets have $M_k=5$ RUs for each UE. Other simulation parameters used in the simulations are given in Table I. Perfect channel estimation is assumed; RF imperfections, hardware impairments, phase noise, power amplifier nonlinearities, etc. are omitted from the scope of this work, which is not expected to affect the qualitative conclusions since it affects equally both unconstrained and routing-constrained performance.

\begin{table}[b!]
\caption{Simulation Model Specifications}
\begin{center}
\centering
\begin{tabular}{|p{3cm}|p{4.5cm}|}
\hline
\textbf{Parameter} & \textbf{Value}\\
\hline
Carrier frequency & 28 GHz\\
\hline
Bandwidth & 200 MHz \\
\hline
Nrof h/v (4x4) subarrays & 2 / 1 (affects only array gain) \\
\hline
Duplexing & TDD \%50 DL \\
\hline
UE / RU height & 1.5 m / 1.5 m \\
\hline
Propagation model & 3GPP InH \cite{R21} \\
\hline

RU power & 13 dBm \\
\hline
UE noise figure & 10 dB \\
\hline
FH segment capacity & 5, 10, 100 (x2 for the corner segments) \\
\hline
Max path length & Equal to FH segment capacity \\
\hline
Serving RU subset size & 5 (for each UE)\\
\hline
Nrof realizations & 50 \\
\hline
\end{tabular}
\end{center}
\end{table}

\subsection{Performance evaluation}
\begin{figure}
 \centering
\begin{subfigure}[t!]{0.44\textwidth}
 \centering
\includegraphics[width=\linewidth]{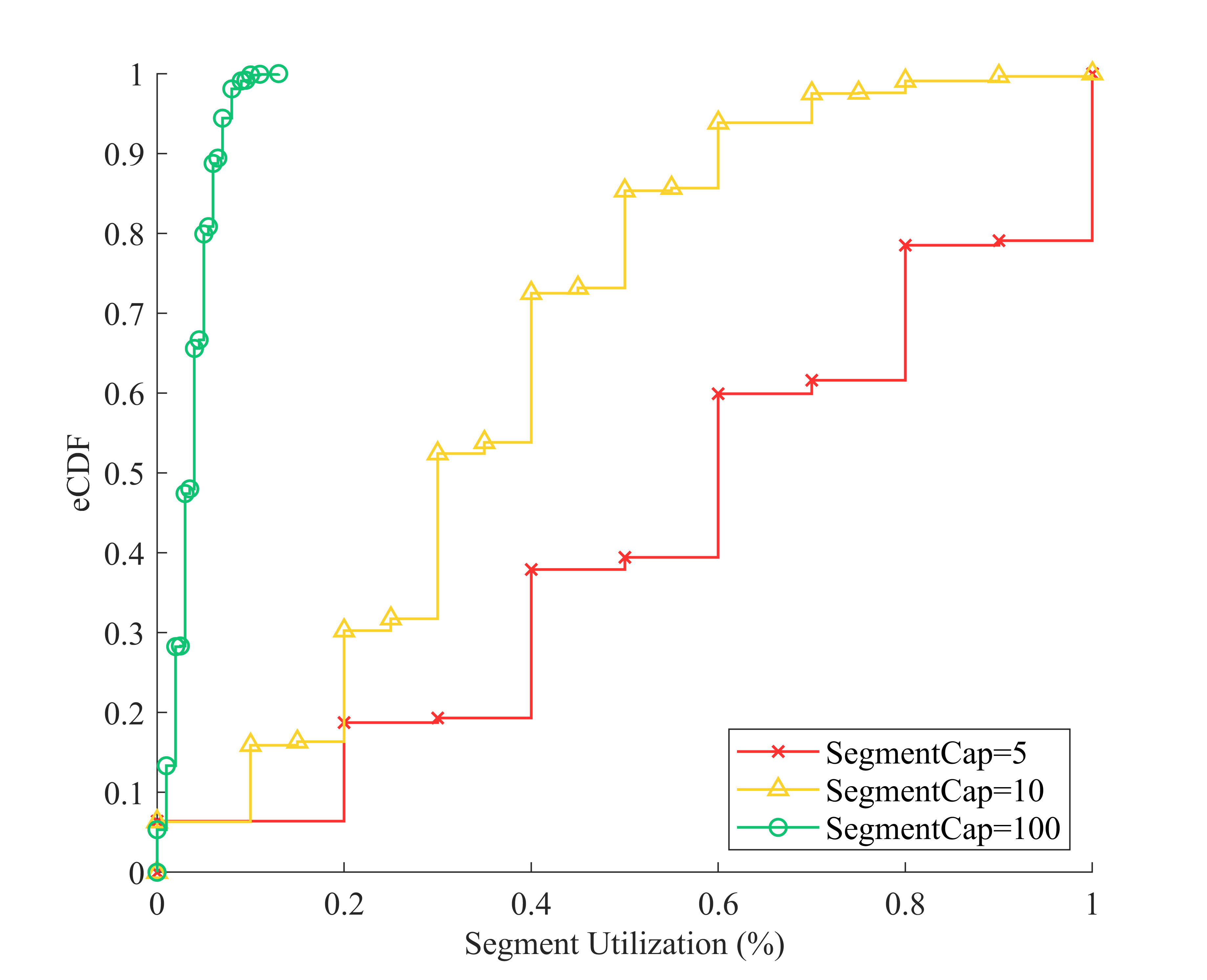} 
\caption{ $K=8$}
\label{fig:3a}
\end{subfigure}
\begin{subfigure}[t!]{0.43\textwidth}
\centering
\includegraphics[width=\linewidth]{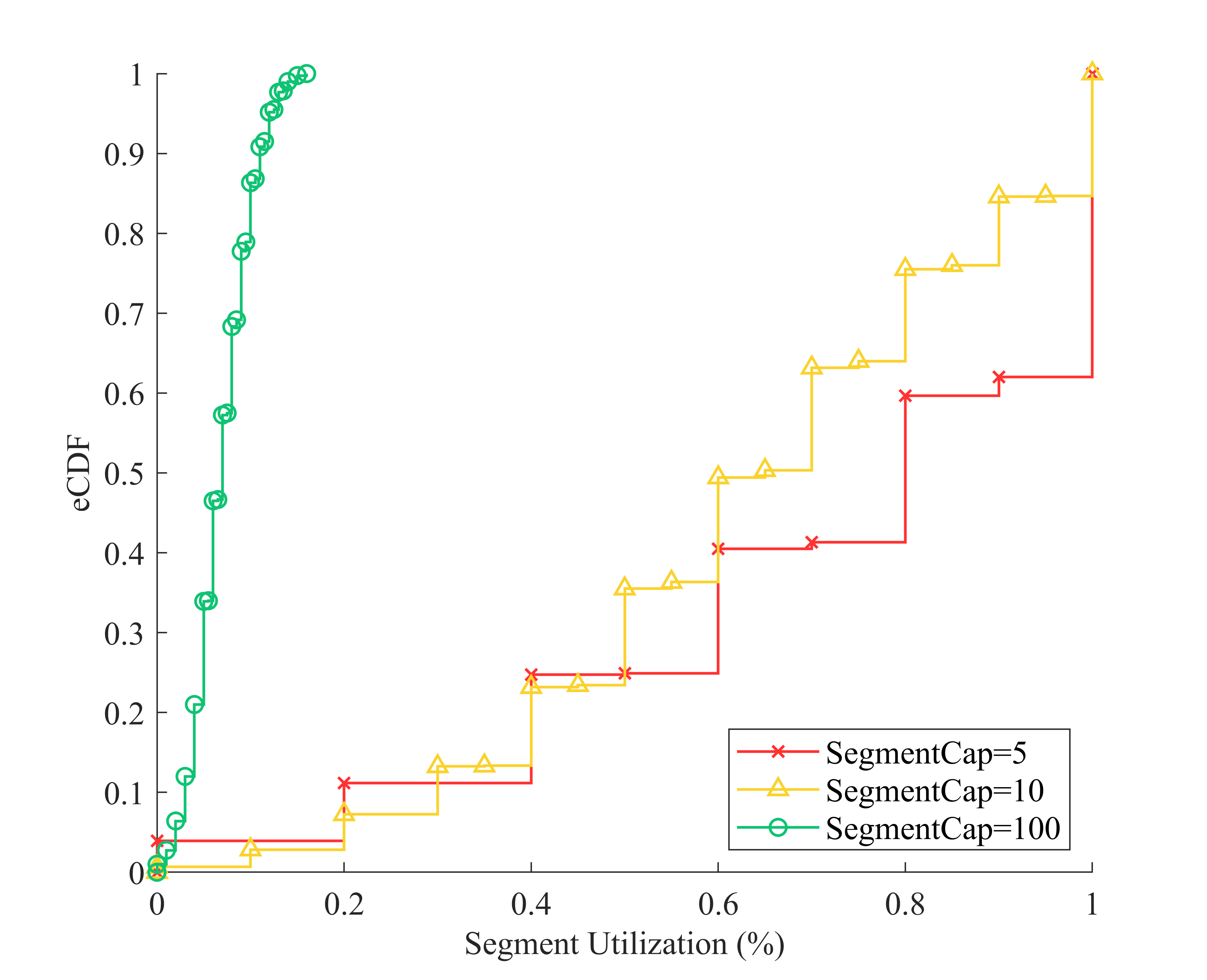}
\caption{$K=15$}
\label{fig:3b}
\end{subfigure}
\centering
\caption{Segment Utilization with $M_k=5, \forall{k}$.}
\label{fig:3}
\end{figure}

\textbf{Segment Utilization}: The Cumulative Distribution Function (CDF) of segment utilization for given segment capacities is shown in Fig. \ref{fig:3}. The results illustrate that, when segment capacity is low, the degree of segment utilization is higher. As segment capacity increases, the utilization reduces. However, as the number of UEs increases, the average segment utilization also increases. When segment capacity is limited, such as 5 packets per time unit, increasing the number of UEs in a case of low utilization would not further enhance the segment utilization due to the fact that some of the UEs are dropped even at 8 UE scenario.

\begin{figure}
 \centering
\begin{subfigure}[t!]{0.43\textwidth}
 \centering
\includegraphics[width=\linewidth]{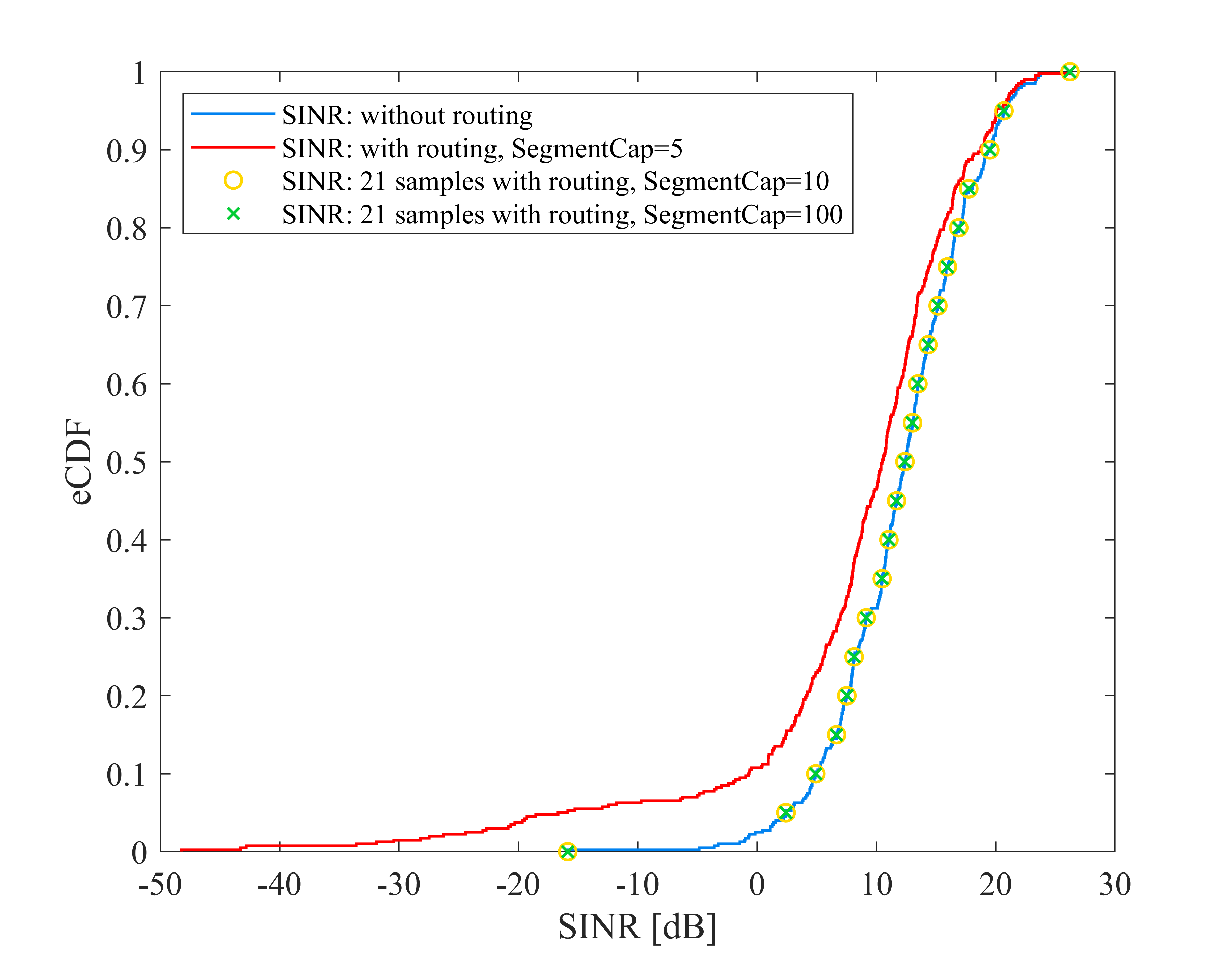} 
\caption{$K=8$}
\label{fig:4a}
\end{subfigure}
\begin{subfigure}[t!]{0.43\textwidth}
\centering
\includegraphics[width=\linewidth]{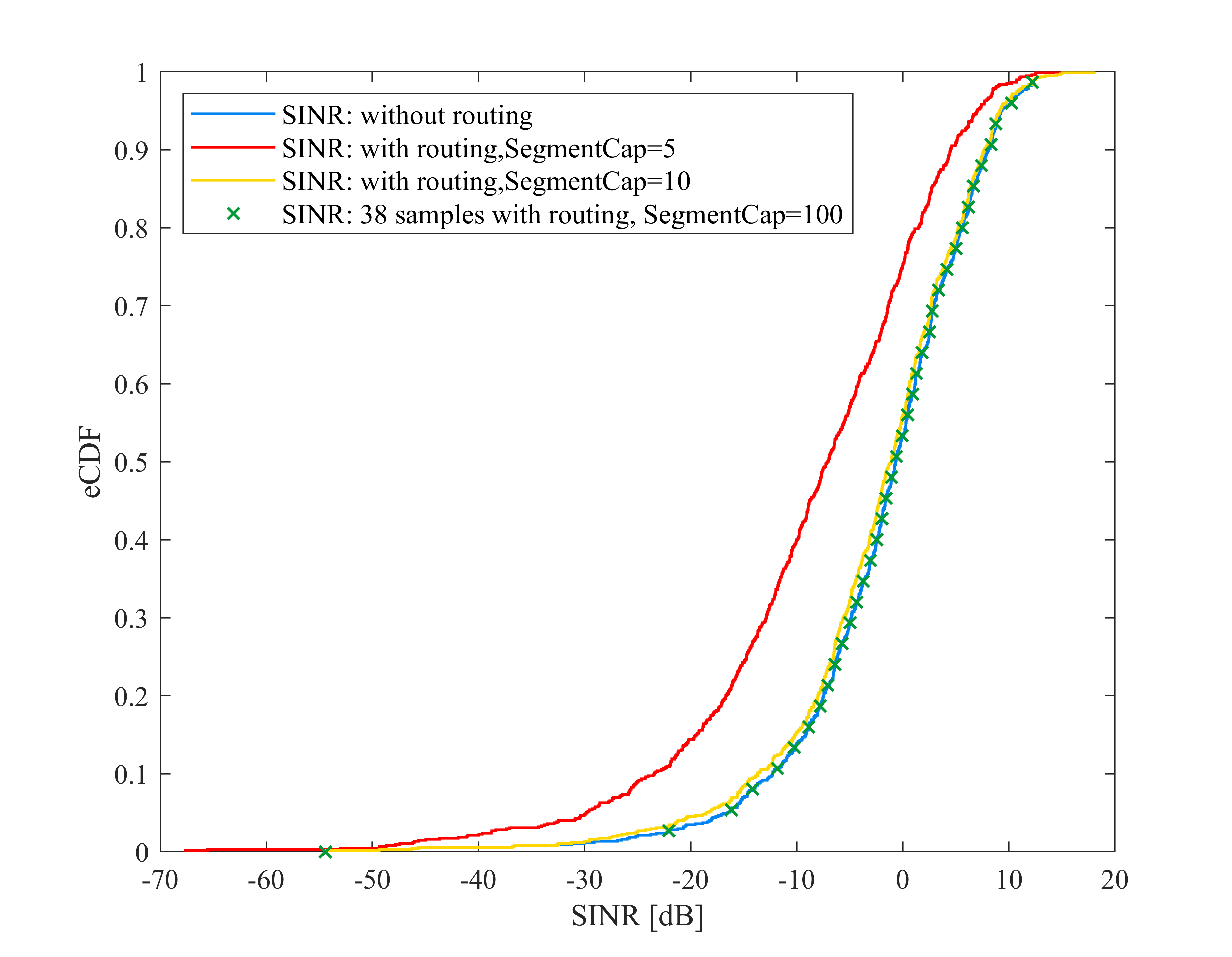}
\caption{$K=15$}
\label{fig:4b}
\end{subfigure}
\centering
\caption{CDF of UE SINRs with $M_k=5, \forall{k}$.}
\label{fig:4}
\end{figure}

\textbf{SINR}: Fig. \ref{fig:4} shows the SINR performance without and with routing impact considering different segment capacities for both $K=8$ (Fig. \ref{fig:4a}) and $K=15$ (Fig. \ref{fig:4b}). Fig. \ref{fig:4b} shows that when the number of UEs is high and more routing failures occur, the resulting SINR is decreased especially for lower segment capacities. It should be noted that when segment capacity is set to 100, the FH capacity is essentially unlimited and yields SINR performance equal to unconstrained routing. Fig. \ref{fig:4b} also shows that when segment capacity is 10, there are limited route discovery failures and consequently some UE drops which slightly affect the SINR.

\begin{figure}
 \centering
\begin{subfigure}[t!]{0.43\textwidth}
 \centering
\includegraphics[width=\linewidth]{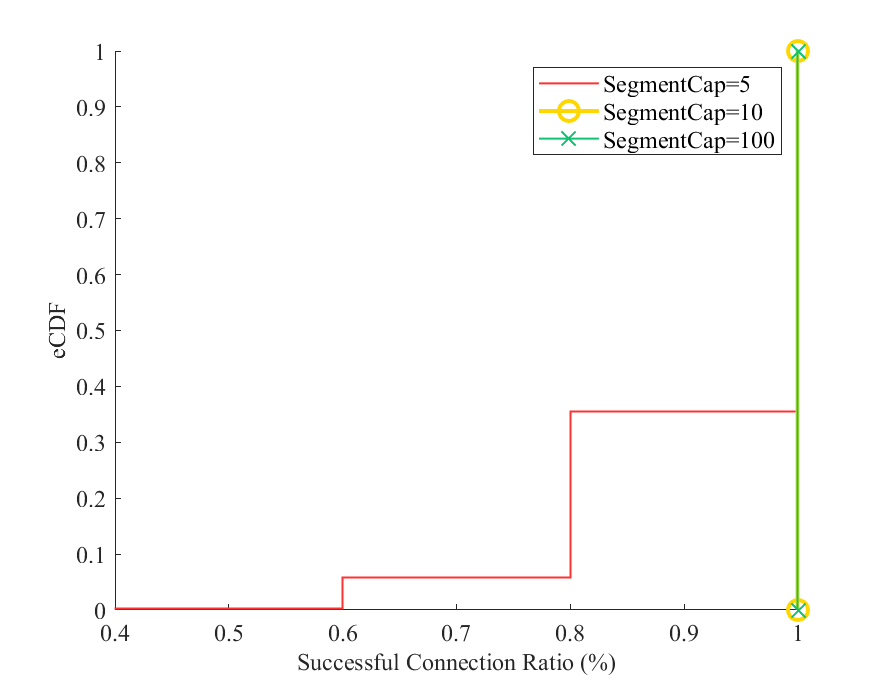} 
\caption{$K=8$}
\label{fig:5a}
\end{subfigure}
\begin{subfigure}[t!]{0.43\textwidth}
\centering
\includegraphics[width=\linewidth]{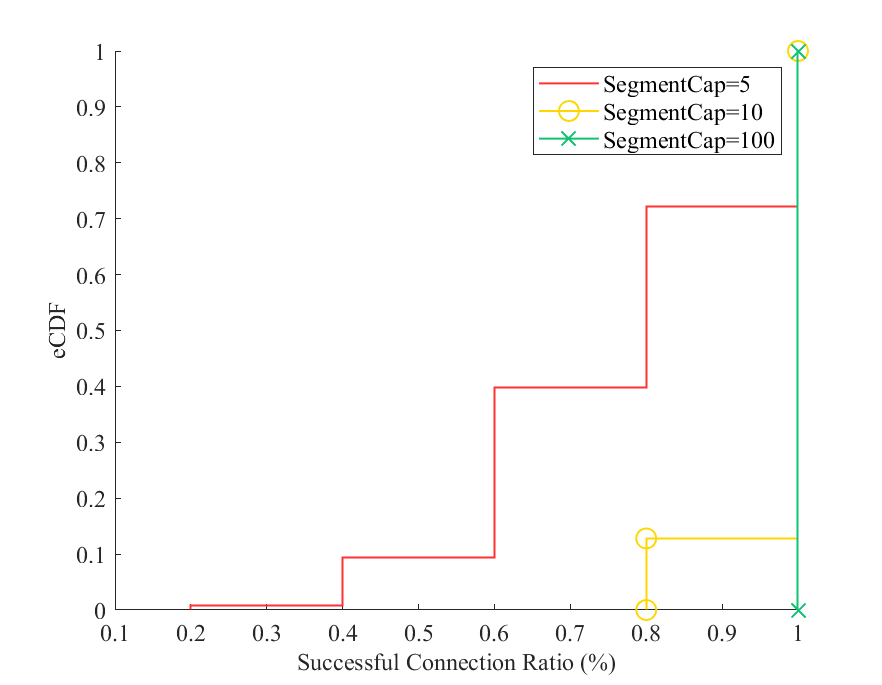}
\caption{$K=15$}
\label{fig:5b}
\end{subfigure}
\centering
\caption{CDF of Successful Connection Ratio for serving RU subsets with $M_k=5, \forall{k}$.}
\label{fig:5}
\end{figure}

\textbf{Successful Connection Ratio}: Fig. \ref{fig:5} shows that the probability of RU connection failure increases as the segment capacity constraint gets tighter. When the segment capacity is limited, some of the segments that are connected to the serving RUs will be occupied even for a lower number of UEs. Fig. \ref{fig:5a} shows that there is a full successful connection ratio for all RUs in each subset of $M_k=5,  \forall{k}$, for a segment capacity 10 or higher. However, when the number of UEs increases from 8 to 15, Fig. \ref{fig:5b}, connections to some of the serving RUs will be failed, hence the results become worse for the segment capacity of 5. Almost each UE is served by at least two RUs, probability of each UE is being served by at most two RUs is around 10\%. For UEs being served by 2-4 connected RUs (the interval for successful connection ratio of 20\%-80\%), the UEs get still service, but routing failures affect the SINR.

\begin{figure}
 \centering
\begin{subfigure}[t!]{0.43\textwidth}
 \centering
\includegraphics[width=\linewidth]{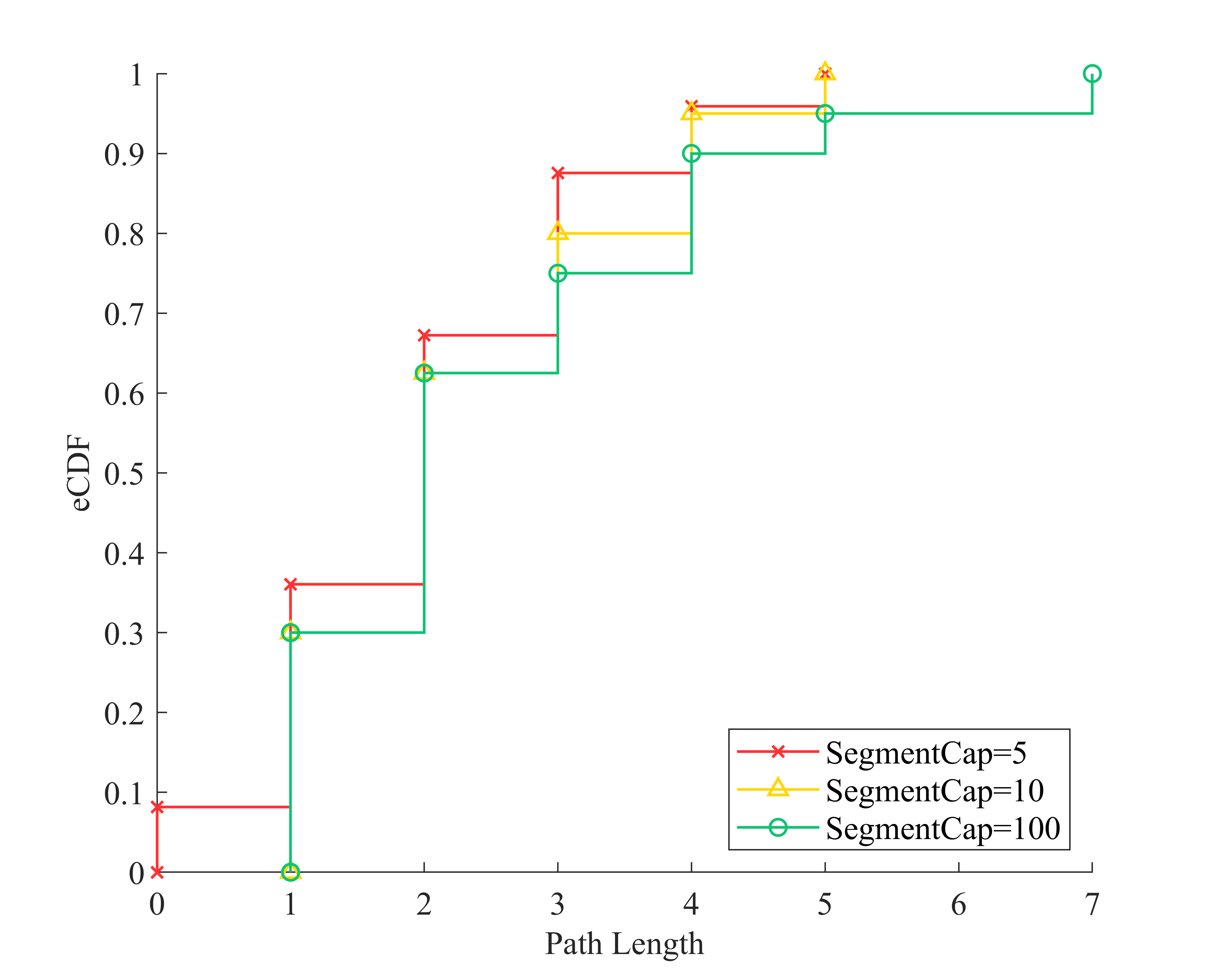} 
\caption{$K=8$}
\label{fig:6a}
\end{subfigure}
\begin{subfigure}[t!]{0.43\textwidth}
\centering
\includegraphics[width=\linewidth]{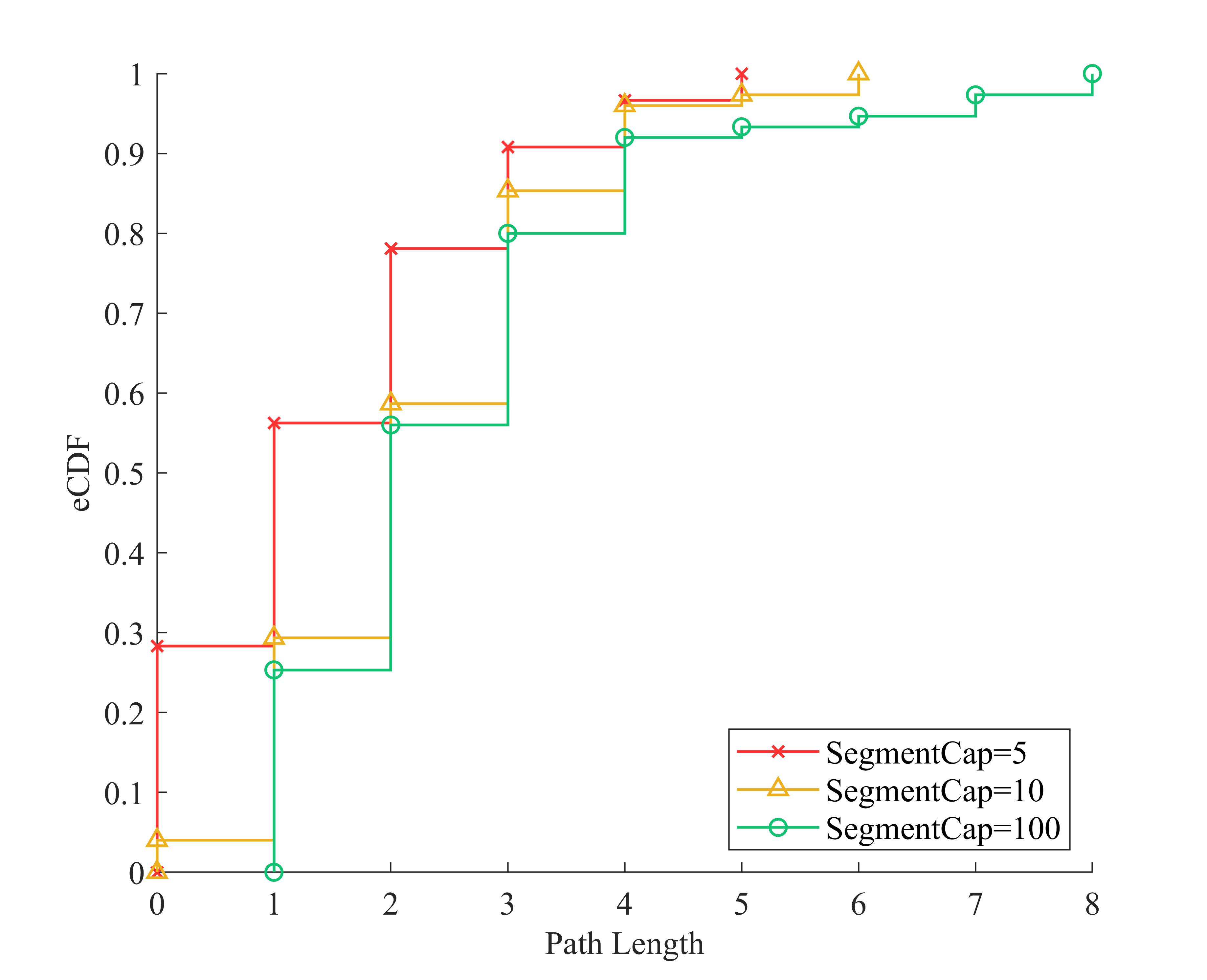}
\caption{$K=15$}
\label{fig:6b}
\end{subfigure}
\centering
\caption{CDF of Path Length from ARU to serving RUs.}
\label{fig:6}
\end{figure}

\textbf{Path Length from ARUs to the serving RUs}: We also studied the length distribution of the established paths from the ARU to the RU in L2 and found that it is affected by the constraints such as segment capacity and the number of UEs. In Fig. \ref{fig:6a}, the maximum path length for segment capacities 5 and 10 is 5 whereas for unlimited segment capacity, e.g., 100, it can be increased up to 7. This indicates that the routing algorithm tries to find alternative paths even with a high path length to avoid UE drop. In Fig. \ref{fig:6b}, the maximum path length is increased to 8 for segment capacity 100. This is directly related to Eq.\ref{Eq5} due to load balancing in individual segment utilization. Whenever the segment capacity is high, the route discovery algorithm calculates more alternative routes with higher path lengths. Therefore, routes with path length 1 are decreased from 30\% to 22\% when $K$ changes from 5 to 10. The algorithm tries to use longer path lengths in a route to avoid repeatedly occupying a specific segment when there are more UEs. For instance, Fig. \ref{fig:6b} shows that there are some routes utilizing 6, 7, or even 8 segments. In delay-sensitive scenarios, it may be preferable to set a lower maximum path length to reduce the number of alternative routes considered from source to destination. On the other hand, that may cause UE drops at lower segment capacities, so to strike a balance in the proposed algorithm the maximum path length could be made to depend on segment capacity. As a result, Fig. \ref{fig:6b} shows that in case of no limitations in path length and segment capacity, the maximum path length that can be considered in future work can be 8 when $K=15$. 

\section{Conclusion and Future Works}
In this study, a two-level routing-based DL transmission is implemented where DU will connect to the ARUs in the first level, and in the second level, the selected ARUs make a connection with the selected serving RUs to the corresponding UEs. The results show that a sufficient segment capacity is a prerequisite to avoid route failures and resulting per-UE SINR reduction, where controlling the UE load and serving RU group size can mitigate such degradation.
The simulation results show that there is a relation between maximum path length and segment capacity usage such that segment capacity and its utilization should be considered for the proper choice of path length.
As a continuation of this work, joint optimization of routing and RU selection, and alternative ARU selection can be investigated.

\section*{Acknowledgment}
This work was supported by The Scientific and Technological Research Council of Turkey (TUBITAK) through the 1515 Frontier Research and Development Laboratories Support Program under Project 5169902 and has been partly funded by the European Commission through the EU Horizon 2020 project REINDEER (grant agreement No. 101013425).

\end{document}